\documentclass[conference]{IEEEtran}

\usepackage{soul,color}
\ifCLASSINFOpdf
   \usepackage[pdftex]{graphicx}
\else
\fi

\usepackage{algorithm}
\PassOptionsToPackage{noend}{algpseudocode}
\usepackage{algpseudocode}

\usepackage{url}

\begin{document}
%
\title{Adaptive QoS of WebRTC for\\Vehicular Media Communications}

\author{\IEEEauthorblockN{\'Angel Mart\'in, Daniel Mej\'ias, Zaloa Fern\'andez,\\Roberto Viola, Josu P\'erez, Mikel Garc\'ia and Gorka Velez}
\IEEEauthorblockA{Vicomtech Foundation\\
Basque Research and Technology Alliance\\
San Sebasti\'an, 20009 Spain\\
Email: amartin@vicomtech.org}
\\
\IEEEauthorblockN{Pablo Angueira}
\IEEEauthorblockA{Department of Communications Engineering\\ 
University of the Basque Country\\
Bilbao, 48013 Spain\\
Email: pablo.angueira@ehu.eus}
\and
\IEEEauthorblockN{Jon Montalb\'an}
\IEEEauthorblockA{Department of Electronic Technology\\ 
University of the Basque Country\\
Bilbao, 48013 Spain\\
Email: jon.montalban@ehu.eus}
}


%


\maketitle

 \IEEEoverridecommandlockouts
 \IEEEpubid{\begin{minipage}{\textwidth}\ \\\\\\\\\\[12pt]\centering
Á. Martín et al., "Adaptive QoS of WebRTC for Vehicular Media Communications", 2022 IEEE International Symposium on Broadband Multimedia Systems and Broadcasting (BMSB), 2022, pp. 1-6, doi: 10.1109/BMSB55706.2022.9828782. \copyright 2022 IEEE.  Personal use of this material is permitted.  Permission from IEEE must be obtained for all other uses, in any current or future media, including reprinting/republishing this material for advertising or promotional purposes, creating new collective works, for resale or redistribution to servers or lists, or reuse of any copyrighted component of this work in other works.
 \end{minipage}}

\begin{abstract}
Vehicles shipping sensors for onboard systems are gaining connectivity. This enables information sharing to realize a more comprehensive understanding of the environment. However, peer communication through public cellular networks brings multiple networking hurdles to address, needing in-network systems to relay communications and connect parties that cannot connect directly.
Web Real-Time Communication (WebRTC) is a good candidate for media streaming across vehicles as it enables low latency communications, while bringing standard protocols to security handshake, discovering public IPs and transverse Network Address Translation (NAT) systems. 
However, the end-to-end Quality of Service (QoS) adaptation in an infrastructure where transmission and reception are decoupled by a relay, needs a mechanism to adapt the video stream to the network capacity efficiently.
To this end, this paper investigates a mechanism to apply changes on resolution, framerate and bitrate by exploiting the Real Time Transport Control Protocol (RTCP) metrics, such as bandwidth and round-trip time. The solution aims to ensure that the receiving onboard system gets relevant information in time.
The impact on end-to-end throughput efficiency and reaction time when applying different approaches to QoS adaptation are analyzed in a real 5G testbed.
\end{abstract}


%
\IEEEpeerreviewmaketitle

\section{Introduction}


Advanced Driver Assistance Systems (ADAS) have turned vehicles into an Internet of Thing (IoT) ecosystem with onboard sensors to understand the driver status, the road environment and the traffic context and are subjected to stringent requirements in terms of reliability and latency \cite{zeadally2020vehicular}.
Their evolution towards a connected swarm, cooperating to expand the individual understanding with sensors from surrounding vehicles and systems, becomes crucial to get a more holistic and broad view. This way, more complex applications such as platooning, cooperative driving or see-through are possible \cite{velez2020}.

However, peer communication between vehicles through public cellular networks brings multiple networking hurdles to address. The problems raised here go from the need to discover the public Internet Protocol (IP) addresses and ports in order to interconnect endpoints to the management of heterogeneous IPv4 or IPv6 addresses or the ability to cross Network Address Translation (NAT) systems \cite{jim2007}.
Accordingly, industrial applications include in-network systems to relay communications and connect parties that cannot connect directly. Here, the relay is employed by parties to remove all the network barriers to communicate with each other.

WebRTC technology is a good option to send video flows between automotive systems \cite{velez2021}. It compiles different standard technologies to bridge peers from different network domains, such as Session Traversal Utilities for NAT (STUN) to negotiate endpoints behind a network infrastructure, Datagram Transport Layer Security (DTLS) to perform the credentials handshake and protect the data flows, and Real Time Transport Control Protocol (RTCP) to capture metrics which monitor network spanning bandwidth and round-trip time.
Moreover, WebRTC can be used with Traversal Using Relay NAT (TURN) to enable relay communications.
Among the available WebRTC solutions, Janus represents a widely employed market solution, as it ships all the mentioned technologies to bridge a production infrastructure \cite{amirante2014janus}.



One essential feature, which is not addressed in these contexts, is the end-to-end adaptation of QoS efficiently. Once a relay comes into play, the RTCP monitoring is limited to the decoupled paths (upload and download) from the peers to the relay server. The adaptation to the more restrictive uplink or downlink path is crucial to ensure that the video stream is adapted to the network capacity of each side.

This paper proposes two solutions for performing the end-to-end QoS adaptation to make the external video source valid in reliability and timing for the onboard system of the vehicle receiving the video. Specifically, this work provides the following relevant contributions:
\begin{itemize}
    \item A mechanism based on bandwidth and round-trip time thresholds to downgrade or upgrade the bitrate, framerate and resolution of the video stream to enforce the reception of a valid stream in terms of reliability and delay.
    \item A first approach, with video transcoding ability at the relay server focusing on the quick reaction to network performance at download path without any additional messaging apart from the standard stack.
    \item A second approach, without transcoding ability forwarding receiver side RTCP values towards the transmitter to gain end-to-end efficiency while adding some latency in the reaction elapsed until the reception of the RTCP report from the receiving peer to adapt the video stream.
    \item A study on the overheads in terms of employed throughput and adaptation latency when comparing the different approaches for delivering onboard cameras on top of a real 5G infrastructure.
\end{itemize}

This paper is structured as follows. First, section \ref{sec:related} focuses on relay solutions for peer communications and QoS adaptation techniques in those infrastructures. Then, section \ref{sec:solution} and section \ref{sec:implementation} present the proposed adaptation of QoS for WebRTC technology and the vehicular setup employed for the tests in a 5G testbed respectively. After, section \ref{sec:results} describes the results to assess the pros and cons of each tested approach. Finally, section \ref{sec:conlusion} sums up the paper topics and outlines some open questions.

\section{Related Work}
\label{sec:related}


QoS adaptation is a core research area in 5G networks which mainly focuses on the allocation or reservation of resources in the radio \cite{ergen2003} or the wired network segments \cite{jawad2018}, and the application of network slicing policies \cite{cominardi2018}. Nevertheless, before they come into play, video streaming protocols and infrastructures need to take care of themselves \cite{shen2018}. 

WebRTC is a technology that supports video, audio and generic data communications between peers. Moreover, the WebRTC stack includes RTCP reports which facilitate information to the origin to adapt the video encoder to the network metrics and guarantee the QoS. Alternatively, DiffServ Code Point (DSCP) can be used to assess end-to-end network measurements and to evaluate QoS policies \cite{barik2018}.

The communication of peers usually needs relay servers to address the limitations and blocking hurdles imposed by network functions of different public networks \cite{jim2007}. Specifically, in video streaming communications, TURN is a widely employed technology with open-source solutions ready to be deployed and used. This technology needs both peers to know the TURN credentials to access it. Janus is a Software as a Service solution ready to be deployed in cloud infrastructures and enabling different communication models as two side communications or multi-party transmission \cite{amirante2019}.

However, both approaches lack end-to-end adaptation mechanisms to track network performance which would entail a video encoding downgrade or upgrade to achieve a reliable and timely information delivery.


\section{QoS Adaptation of WebRTC sessions in mobile networks}
\label{sec:solution}

\begin{figure*}[htp]
\centering
\includegraphics[width=0.95\textwidth,clip,keepaspectratio]{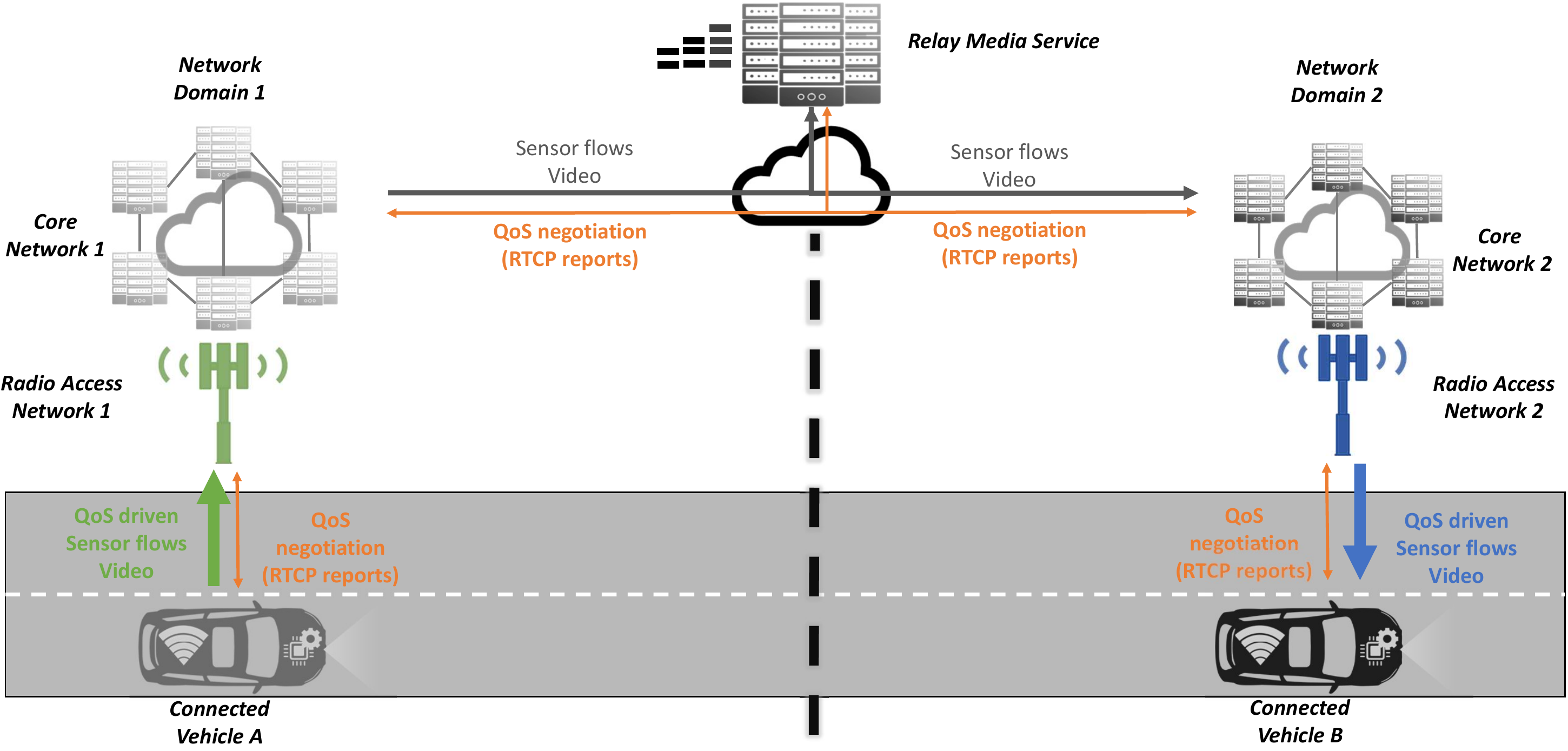}
\caption{QoS-driven vehicular media communications with relay media service.}
\label{fig:relay}
\vspace{-0.5cm}
\end{figure*}


In order to identify the context where a mechanism to deal with QoS adaptation of a WebRTC session, it is important to underline the potential communication options when two vehicles communicate with each other through mobile networks:
\begin{itemize}
    \item \textit{Direct Communication is possible}. This scenario implies that both vehicles are subscribed to the same cell and the same Public Land Mobile Network (PLMN) id or that their respective cellular operator enables the communication (sockets) between User Equipments (UEs). This is not common, and usually, the Mobile Network Operator (MNO) does not allow communication between UEs when they have not contracted public IPs. In this scenario, the network stats reported by RTCP can monitor the full communication path and identify any issue related to the uplink, wired backhaul/core and downlink. This way, the sender of a WebRTC media session could upgrade or downgrade the media to better fit into the available network performance.
    \item \textit{Direct Communication is not possible}. This scenario implies a gateway, ideally at the network edge, as a service provided by a Multi-access Edge Computing (MEC) infrastructure to minimize the traffic latency and save traffic at the core. This gateway establishes connections with peers and solves the previously mentioned common limitation, as the UEs open a socket with a server and do not communicate directly. In this case, the connection monitoring is split, as the peers are isolated and the sender only gets feedback of the uplink and the path to the gateway (upload). Moreover, the gateway gets feedback of the path to the Radio Access Network (RAN) and the downlink (download). Here, the sender, which performs the heavy task of media encoding, only has visibility of the upload connection. Thus, any bottleneck present at the receiver for download connection will not produce a reaction.
\end{itemize}

In order to address the limitation of monitoring and adaptation when direct communication is not possible, a mechanism to adapt the video encoding setup to the different paths present when a gateway acts as a relay server to communicate peers, as shown in Figure \ref{fig:relay}, is required.

To be able to provide adaptation to any potential connectivity issue in the connection path, two different approaches are proposed:
\begin{itemize}
    \item \textit{Transcoding Relay:} the first approach places transcoding abilities at the relay server. This is able to parse and process the RTCP reports of the download path echoed by the reception peer. This option implies the capacity to perform media transcoding with live settings from the RTCP values. This option does not scale well as the transcoding means heavy processing. Moreover, the sender could incur some overheads sending a bitrate, framerate and resolution that would be downgraded at the relay, wasting processing and computing resources. Furthermore, usually, the MEC infrastructures are not offered shipping GPU features. 
    \item \textit{Reporting Relay:} in the second approach, the relay server forwards the RTCP reports generated between the relay and the receiver towards the sender peer through a WebRTC data channel. This way, the relay provides awareness of the receiver download metrics to the sender. Then, the sender can decide to adapt the encoding settings both to the upload path to the gateway and to the receiver download path capacity. This solution brings scalability, forwarding media flows and messages from data channel without significant data overheads. 
\end{itemize}

\begin{figure}[htp]
\centering
\includegraphics[width=0.5\textwidth,clip,keepaspectratio]{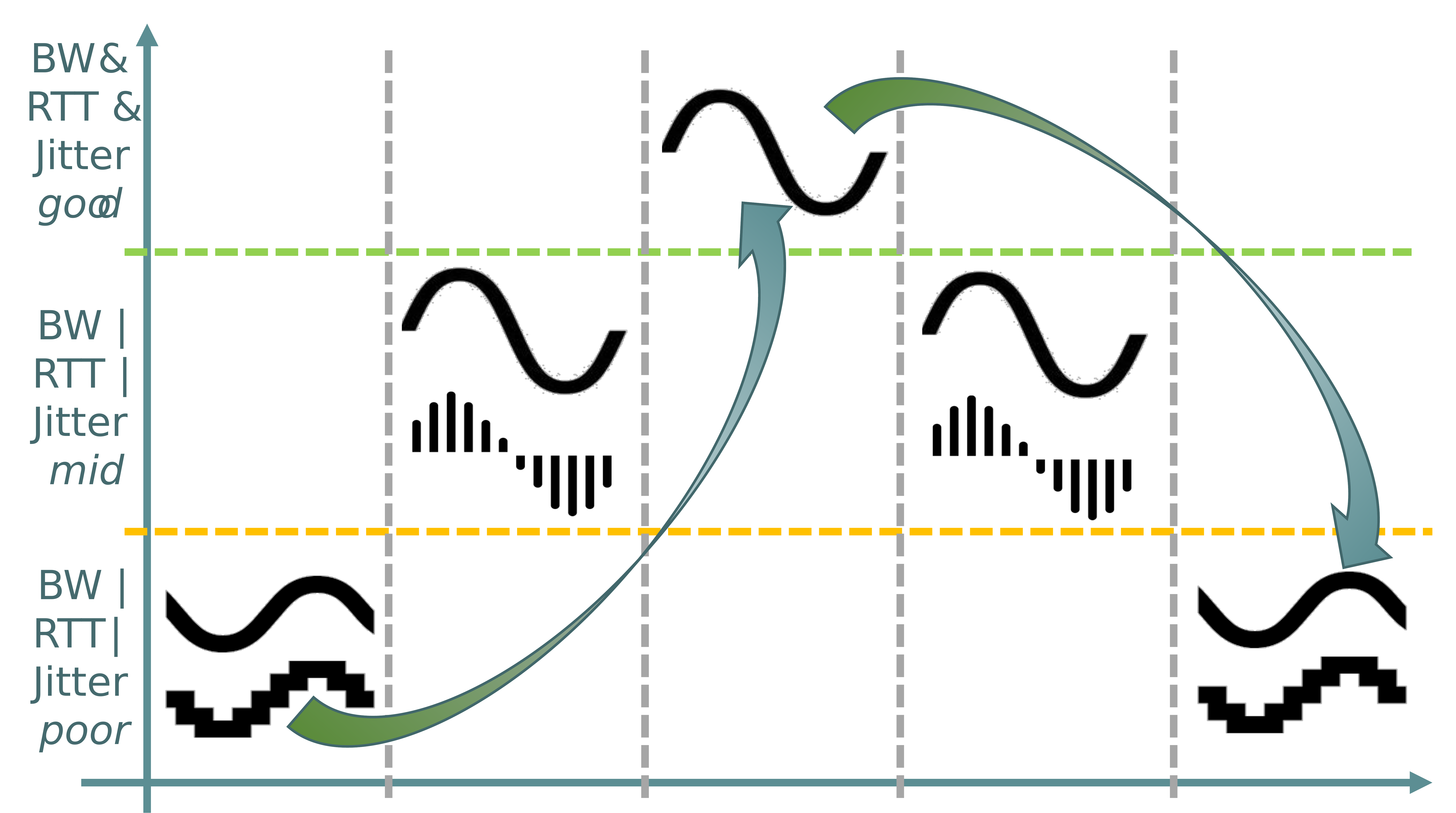}
\caption{3-level rate control adaptation based on Bandwidth, Round trip-time and Jitter metrics.}
\label{fig:levels}
\vspace{-0.5cm}
\end{figure}

This mechanism runs at the gateway and the sender for the transcoding relay or just at the sender for the reporting relay is as follows. The adaptation mechanism allows for adjusting the resolution, framerate and bitrate of the video according to thresholds on the assessed bandwidth and round-trip time of the upload and download paths from the peers to the relay server, as depicted in Figure \ref{fig:levels}.

The Algorithm \ref{alg:algorithm} describes in detail the bitrate adaptation mechanism executed at the sender. Accordingly, 3 different levels are considered. The nominal bitrate, resolution and framerate are employed when the bandwidth, round-trip-time (RTT) and jitter values are good enough (depending on the bitrate). The bitrate is downgraded when some of the bandwidth, the RTT, or the jitter are not optimal. In this mid-level, to keep the video fidelity, the video is subsampled with a lower framerate and shorter GOP to keep the ability to quickly access to a video stream, which is highly important for spontaneous vehicular communications. Finally, when the network conditions are unfavorable and the values of the bandwidth, the RTT, or the jitter are poor, the bitrate goes to a minimum, where the framerate is again downgraded and the resolution, meaning \textit{Width x Height} (WxH), is also reduced. As the framerate, in this case, is minimum, the GOP size is needed to be bigger, meaning slower access to the video stream. Otherwise, the encoder would use too many key frames damaging the compression rate. This way, we try to provide a continuous video signal to onboard computer vision systems to at least receive a video signal even with reduced resolution or framerate, trying to keep captured information and image details.

\begin{algorithm}
\renewcommand{\algorithmicrequire}{\textbf{Input:}}
\renewcommand{\algorithmicensure}{\textbf{Output:}}
\caption{3-levels Adaptive Rate Control.}
\label{alg:algorithm}
\begin{algorithmic}
\Function{adaptiveRate}{bw$_{t}$, rtt$_{t}$, jit$_{t}$} \Comment{\parbox[t]{0.23\linewidth}{when a RTCP report comes}}
\Require bw$_t$ \Comment{measured bandwidth}
\Require rtt$_t$ \Comment{measured delay}
\Require jit$_t$ \Comment{measured jitter}
\Ensure bitrate$_{t+1}$ \Comment{next bitrate}
\Ensure framerate$_{t+1}$ \Comment{next framerate}
\Ensure resolution$_{t+1}$ \Comment{next resolution}
\State $m$ $\leftarrow$ mid divider coefficient \Comment{subsampling for mid level}
\State $p$ $\leftarrow$ poor divider coefficient \Comment{subsampling for poor level}
\If {(bw$_t$, rtt$_t$, jit$_t$) $>$ (bw$^{good}$, rtt$^{good}$, jit$^{good}$)} 
\State bitrate$_{t+1}$ $\leftarrow$ bitrate$^{max}$ \Comment{nominal bitrate}
\State framerate$_{t+1}$ $\leftarrow$ framerate$^{max}$ \Comment{nominal framerate}
\State resolution$_{t+1}$ $\leftarrow$ resolution$^{max}$ \Comment{nominal WxH}
\State gopSize$_{t+1}$ $\leftarrow$ gopSize$^min$ \Comment{quickest access}
\ElsIf {(bw$_t$, rtt$_t$, jit$_t$) $<$ (bw$^{poor}$, rtt$^{poor}$, jit$^{poor}$)} 
\State bitrate$_{t+1}$ $\leftarrow$ bitrate$^{max}$/p \Comment{details kept}
\State framerate$_{t+1}$ $\leftarrow$ framerate$^{max}$/p \Comment{irregular}
\State resolution$_{t+1}$ $\leftarrow$ resolution$^{max}$/p \Comment{W/(p/2)xH/(p/2)}
\State gopSize$_{t+1}$ $\leftarrow$ 1 \Comment{access limited to every second}
\Else
\State bitrate$_{t+1}$ $\leftarrow$ bitrate$^{max}$/m \Comment{mid bitrate}
\State framerate$_{t+1}$ $\leftarrow$ framerate$^{max}$/m \Comment{not so smooth}
\State resolution$_{t+1}$ $\leftarrow$ resolution$^{max}$ \Comment{nominal WxH}
\State gopSize$_{t+1}$ $\leftarrow$ gopSize$^{min}$/m \Comment{still quick access}
\EndIf
\EndFunction
\end{algorithmic}
\end{algorithm}

\section{Implementation}
\label{sec:implementation}

For the implementation of the different actors, different alternatives and frameworks have been employed:
\begin{itemize}
    \item Sender and Receiver. On top of Gstreamer framework for media processing, some applications have been developed  to send and receive video streams in H.264/Advanced Video Codec format (AVC) format. In order to have a fast encoding, NVIDIA plugins are used to employ hardware acceleration and remove any latency from the encoding perspective. Both Gstreamer applications use Interactive Connectivity Establishment (ICE) protocol combined with a STUN server from google to allow the discovery of public IPs even when they are behind a NAT. In terms of WebRTC, the period for the RTCP reports can be configured. Additionally, they are synchronised through an NTP server and include a watermarking mechanism that stamps microsecond accuracy clock in the image, which can be recovered by the receiver to assess the end-to-end latency. The sender includes the ability to restart the GOP, sending a key frame and all the required headers, i.e., Sequence Parameter Set (SQS) and Picture Parameter Set (PPS) in H.264, to immediately apply any change on resolution or framerate.
    \item Transcoding relay. This simple Python gateway enables the communication of peers when they can communicate directly or when they cannot. For the latter, a WebRTC receiver is instantiated to receive the video stream at the server side (upload) and a WebRTC sender is instantiated to deliver the video stream to the receiver (download). The communication between the receiver and sender running in the relay is done directly inside the server through UDP to minimize the end-to-end latency, as the WebRTC signalling takes some time ($\sim$ 500 ms) while internal UDP is instant without signaling. The sender in the relay is able to transcode the video based on applicable RTCP reports (download).
    \item Reporting relay. This Janus gateway enables peers' communication when they cannot communicate directly. Janus, using video rooms and text rooms, is able to forward the sending media in a one to many or many to many manners, which would expand the suitable use cases significantly in an efficient way. Thus, the gateway is able to forward any message as far as RTCP reports of the download are sent to the sender, which by default only has upload visibility. 
\end{itemize}

\section{Results}
\label{sec:results}

All the results have been captured from a setup where the UEs are Quectel RM500Q modems accessing a 5G SA base station operated by an Amarisoft Callbox Pro in the band n78 (3,5 GHz) with 100 MHz of bandwidth which provides more than 200 Mbps of downlink and 120 Mbps uplink with around 20 ms latency. The UEs are connected with USB to two laptops providing network connectivity. The laptops include 16GB RAM, i7-10th gen and NVIDIA graphics card capable of encoding 4K-video in H.264. For the MEC hosting of the Dockers of the gateways, an equivalent machine is employed. This is directly connected to the Amarisoft network Core to avoid any impact there. This way, we employ a real 5G experimentation setup.

To stress good, mid and poor network conditions in the upload or download, we apply some rules in the sender or receiver via the \textit{tc} Linux command, which allows us to add bandwidth limits or to include an artificial latency in the communications of a machine. In this regard, the Table summarises the applied conditions. A script changes every 20 s the bandwidth or latency from the worst value to the best and then goes back to the worst. The Table \ref{tab:bws} represents the limits applied for bandwidth experiments while Table \ref{tab:lats} runs along the added latency. According to some tests done with a ping and an IPERF, the application of such limits to any ongoing or new connection is immediate.

\begin{table}[t]
\caption{Set of limits employed in the Bandwidth experiments.}
\centering
\bgroup
\def\arraystretch{1.2}
\setlength\tabcolsep{2.5pt} 
\label{tab:bws}
\begin{tabular}{|c|c|}
\hline
\textbf{Period} & \textbf{bandwidth (Mbps)} \\
\hline
0-20 & 1 \\
20-40 & 10 \\
40-60 & 100 \\
60-80 & 10 \\
80-100 & 1 \\
\hline
\end{tabular}
\egroup
\end{table}

\begin{table}[t]
\caption{Set of overheads added in the Latency experiments.}
\centering
\bgroup
\def\arraystretch{1.2}
\setlength\tabcolsep{2.5pt} 
\label{tab:lats}
\begin{tabular}{|c|c|}
\hline
\textbf{Period} & \textbf{latency (ms)} \\
\hline
0-20 & 600 \\
20-40 & 100 \\
40-60 & 10 \\
60-80 & 100 \\
80-100 & 600 \\
\hline
\end{tabular}
\egroup
\end{table}

Furthermore, the employed levels for the adaptation are compiled in Table \ref{tab:levels} as they are valid for computer vision systems that identify bounding boxes, supporting Advanced driver-assistance systems (ADAS) applications such as see-through.

\begin{table}[t]
\caption{Encoding parameters for each level.}
\centering
\bgroup
\def\arraystretch{1.2}
\setlength\tabcolsep{2.5pt} 
\label{tab:levels}
\begin{tabular}{|c|c|c|c|c|}
\hline
\textbf{Level} & \textbf{Bitrate (kbps)} & \textbf{Framerate (fps)} & \textbf{Resolution (WxH)} & \textbf{GOP (frames)} \\
\hline
Good & 4000 & 30 & 1920x1080 & 5 \\
Mid & 2200 & 15 & 1920x1080 & 7 \\
Poor & 700 & 5 & 640x360 & 5 \\
\hline
\end{tabular}
\egroup
\end{table}

\begin{figure}[htp]
\centering
\includegraphics[width=0.5\textwidth,clip,keepaspectratio]{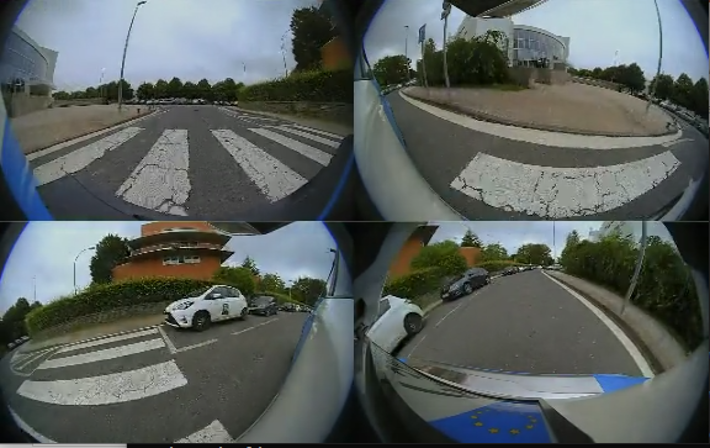}
\caption{Sample of recording employed.}
\label{fig:cams}
\end{figure}

This section contains the analysis of the outcomes obtained from the implementation. In this regard, we have performed 4 different tests running 2000 s of video streaming of a recording captured by onboard cameras on a sensorised vehicle as shown in Figure \ref{fig:cams}. Here the 5 different levels of bandwidth limitations and latency aggregations from Tables \ref{tab:bws} and \ref{tab:lats} have been applied, resulting in 100 changes applied in upload or download.

The first experiment tries to find the impact on the RTCP reporting period in the quicker scenario when the peers can communicate directly and the RTCP reports monitor both upload and download. This means that the sender can make a decision with all the information in hand. In this case, we have tested configuring RTCP reporting period to 500 ms and 1 s. Table \ref{tab:report} shows the better average and standard deviation of the different tests done when applying bandwidth limits and adding latency on the sender or the receiver side.

\begin{table}[t]
\caption{Reporting period results when applying networking issues.}
\centering
\bgroup
\def\arraystretch{1.2}
\setlength\tabcolsep{2.5pt} 
\label{tab:report}
\begin{tabular}{|c|c|c|}
\hline
\textbf{Report Period} & \textbf{Average Sender} & \textbf{Standard Deviation of} \\
\textbf{(ms)} & \textbf{reaction (s)} & \textbf{Sender reaction (s)} \\
\hline
500 & 3.5 & 1.2 \\
1000 & 4.5 & 1.4 \\
\hline
\end{tabular}
\egroup
\end{table}

The results show that the more frequent the RTCP reports are, the quicker the sender becomes aware of that situation through RTCP reports and can react. Thus, we use 500 ms for the RTCP reporting period in the following experiments,.

These experiments also allow us to find the appropriate thresholds for the cellular experimentation setup to trigger the encoding level change. These are compiled in the Table \ref{tab:thresholds}. These values are not optimised to bridge a paramount experience but to check that the mechanisms properly change between the levels when the limitations in bandwidth or latency are applied on the sender or the receiver side.

\begin{table}[t]
\caption{Thresholds for captured RTCP metrics.}
\centering
\bgroup
\def\arraystretch{1.2}
\setlength\tabcolsep{2.5pt} 
\label{tab:thresholds}
\begin{tabular}{|c|c|c|c|c|}
\hline
\textbf{Border} & \textbf{Bandwidth (kbps)} & \textbf{RTT (ms)} & \textbf{Jitter (ms)} \\
\hline
Good/Mid & 10000 & 90 & 2 \\
Mid/Poor & 5000 & 180 & 8 \\
\hline
\end{tabular}
\egroup
\end{table}

With all the parameters configured to be common to all the tests, the three experiments are done are:
\begin{itemize}
    \item Direct communication, where sender and receiver directly exchange RTCP reports.
    \item Simple Python gateway, where all the traffic goes through the gateway and the sender is only able to monitor upload path and download issues are not managed, as the transcoding in the gateway is not scalable.
    \item Janus gateway, where all the traffic goes through the Janus gateway able to provide media delivery to the receiver and forward the receiver RTCP reports to the sender to efficiently apply changes on encoding settings for upload and download issues.
\end{itemize}

Table \ref{tab:results} compiles all the results for the three different setups. Here, the time of the sender to get the awareness of the issues from the RTCP reports and the time elapsed from that moment to the moment the receiver gets updated video settings are compiled.

\begin{table}[t]
\caption{Reaction time at sender and elapsed time at receiver side when applying networking issues.}
\centering
\bgroup
\def\arraystretch{1.2}
\setlength\tabcolsep{2.5pt} 
\label{tab:results}
\begin{tabular}{|c|c|c|}
\hline
\textbf{Setup} & \textbf{Average Sender} & \textbf{Average Receiver} \\
\textbf{} & \textbf{reaction (s)} & \textbf{encoding update (s)} \\
\hline
Direct Communication & 2.5 & 0.7 \\
Simple Gateway & 3.3 & 0.7 \\
Janus Gateway & 3.6 & 1.0 \\
\hline
\end{tabular}
\egroup
\end{table}

From the Table \ref{tab:results}, it becomes evident several aspects. First, the RTCP reports take a lot of time to sense the changes in the network performance. Second, direct communication is the best case but not always viable, as noticed in commercial Long Term Evolution (LTE) and 5G networks. The gateways are able to do the work, adding some latency to the sender's reaction as the messages need to travel from the receiver to the sender via the gateway. Third, our mechanism takes from 700 ms to 1 s to apply the new encoder settings from the moment the sender decides to change it to the moment the received video stream includes such changes. Last but not least, in the Simple Gateway scenario, as the sender has no visibility of the download's metrics, we have just applied changes in the upload, as this solution is not able to adapt to issues raised in the download.

\section{Conclusions and Future Work}
\label{sec:conlusion}


This paper proposes a mechanism based on network stats reported by RTCP to adapt the encoder parameters of video streams for mobile communications in the vehicular domain. This mechanism enables the adaptation efficiently providing networking metrics to the sender in a standard manner. This is highly beneficial for scenarios where the vehicles cannot communicate directly with each other, which is a common situation in commercial networks.

The solution is tested in a real 5G network where artificial bottlenecks are added to stress the upload and download paths. In the future, we plan to use an outdoor setup to perform realistic experiments, including cars and real ADAS applications to test the reliability of networks for such applications and optimize the reaction times.


\section*{Acknowledgment}
This research was supported by the European Union’s Horizon 2020 research and innovation programme under grant agreement No. 957360 (5GMETA project) and the Spanish Centre for the Development of Industrial Technology (CDTI) and the Ministry of Economy, Industry and Competitiveness under grant/project \textit{CER-20191015 / Open, Virtualized Technology Demonstrators for Smart Networks (Open-VERSO)}.




%


\bibliographystyle{IEEEtran}
\bibliography{main.bib}

\end{document}